# Physical Removal of Metallic Carbon Nanotubes from Nanotube Network Devices Using a Thermal and Fluidic Process

*Alexandra C. Ford,[1] Michael Shaughnessy,[1] Bryan M. Wong,[1] Alexander A. Kane,[1] Oleksandr V. Kuznetsov,[2] Karen L. Krafcik,[1] W. E. Billups,[2] Robert H. Hauge,[2] François Léonard[1]\**

[1]Sandia National Laboratories, Livermore, California 94551, United States

[2]Department of Chemistry, Rice University, Houston, Texas 77005, United States

*fleonar@sandia.gov

**Abstract -** Electronic and optoelectronic devices based on thin films of carbon nanotubes are currently limited by the presence of metallic nanotubes. Here we present a novel approach based on nanotube alkyl functionalization to physically remove the metallic nanotubes from such network devices. The process relies on preferential thermal desorption of the alkyls from the semiconducting nanotubes and the subsequent dissolution and selective removal of the metallic nanotubes in chloroform. The approach is versatile and is applied to devices post-fabrication.

# 1. Introduction

For two decades, carbon nanotubes (CNTs) have been explored for a variety of electronic and optoelectronic applications [1]. Recently, devices based on thin films of CNTs [2] have received much interest due to their higher currents and larger light collection or emission areas compared to those based on individual CNTs. However, current CNT thin film devices are composed of both semiconducting and metallic CNTs due to the inability to selectively synthesize CNTs of a given electronic type. Because most electronic and optoelectronic applications rely on semiconducting behavior, the presence of metallic CNTs hinders CNT device development, and physically eliminating or quenching metallic CNTs from networks or arrays remains an important challenge.

Previous efforts in this direction have focused on solution-based separation of nanotubes by electronic type prior to device fabrication [3, 4, 5, 6, 7] or processing of the CNT network after device fabrication. These post-fabrication approaches include Joule heating to burn the metallic CNTs [8], the conversion of metallic CNTs to semiconductors using light irradiation [9], electron beam irradiation [10], and plasma treatment [11, 12], or CNT functionalization with a diazonium reagent to suppress the conductivity of the metallic CNTs [13, 14, 15, 16]. These approaches often lead to degradation of the device performance due, for example, to the introduction of defects in the remaining CNTs.

In this work, we present an approach where the metallic CNTs are physically removed from the network [17] without degrading the device performance. This is accomplished by utilizing recent advances in CNT surface functionalization [18, 19, 20] to demonstrate a method to selectively and physically remove metallic CNTs directly from network devices. The method relies on the preferential thermal desorption of alkyl groups from semiconducting CNTs, and the good solubility of the remaining functionalized metallic CNTs in chloroform, allowing them to be selectively removed. The method is shown to be applicable to devices after fabrication on wafers, including in the presence of electrical contacts. The experimental results are supported by modeling of the nanotube functionalization, the impact on the CNT conductance, and the thermal desorption.

## 2. Experimental details

### 2.1. CNT functionalization

The CNTs used in this work were grown using the HiPCo process using low catalyst amounts, resulting in material containing 2/3 semiconducting CNTs and 1/3 metallic CNTs. Functionalization reactions were achieved as described previously [18, 19]. In brief, addition of either lithium or sodium to a dispersion of CNT bundles in liquid ammonia leads to a debundled CNT salt that serves as a source of electrons. Then, alkyl iodide is reacted with the CNT salt to yield radical anions. The transient radical anions dissociate rapidly to provide dodecyl radicals that form covalent bonds to the sidewalls of the CNT. These reactions are outlined below:

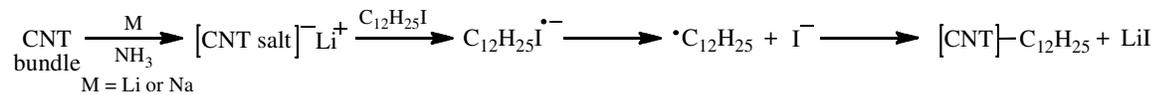

### 2.2. Device fabrication

Once functionalized, the CNTs were suspended in hexane, and then dispersed in chloroform by bath sonication for 10min. CNT network devices were fabricated by spincasting the CNTs dispersed in chloroform onto 90nm SiO$_2$/Si substrates with pre-patterned electrodes at 2000rpm for 30s. The electrodes of spacing ~1μm and width 20μm were patterned using photolithography with a bilayer of liftoff resist and Shipley 1805 photoresist. The contact metal was deposited using electron-beam evaporation of 5nm Ti/35 nm Pd. The dispersed CNTs ranged from ~100nm to 3.5µm in length, with an average length of ~400nm. The devices were annealed in vacuum (~8x10$^{-6}$ torr) at different temperatures between 75 and 585 °C, with the temperature held constant for 20 min at each temperature. Electrical measurements were taken at room temperature in air the next day following each anneal step. Devices treated with chloroform were first annealed to the desired temperature, and then placed in a chloroform bath for ~45s with gentle agitation, followed by additional chloroform rinsing.

*2.3. Atomic force microscopy and image analysis*

Atomic force microscopy (AFM) height and amplitude images were taken using a Digital Instruments Dimension 3000 AFM (Veeco). For statistical analysis of the networks 10µm x 10 µm scans were acquired, and the height images were flattened and corrected for periodic noise from the scan lines. SIMAGIS image analysis software was used to measure the CNT density and length distribution from the combined height and amplitude images using the CNT/Nanotechnology module.

*2.4. Ab inito calculations*

We used a dispersion-corrected PBE-D functional, which is designed specifically for binding energies. Structural optimizations for all alkylated systems were carried out with spin-polarized, projected-augmented-wave potentials in conjunction with a 350 eV plane-wave cutoff. Since the alkyl molecules are over 6 times longer than both the (6,6) and (10,0) CNT unit cells, a large supercell of 17.1 Å along the nanotube axis was chosen to allow separation between adjacent alkyl groups and to match the coverage extracted from analysis of the TGA data as discussed below. To prevent spurious interactions between the alkylated CNTs themselves (which arise from 3D periodic boundary conditions), all structural optimizations were carried out with 50 Å of distance between the walls of neighboring CNTs. Due to the large cell, a k-point sampling of 12 x 1 x 1 was used for the Brillouin zone sampling.

To compute the CNT conductance from the DFT calculations, the method from Ref. 21 was utilized [21]. The bandstructure is computed on a fine 32x1x1 mesh in the Brillouin zone along the direction corresponding to the CNT axis. Reduced supercell models with one alkyl group per unit cell (shown in the atomic model insets of Figure 4) are used to make the calculation computationally tractable. The conductance is calculated from $G(E) = (2e^2/h)T(E)M(E)$ where $T(E)$ and $M(E)$ are the transmission probability and number of modes at energy $E$. The number of modes is obtained from the DFT bandstructure. We consider the ballistic transport regime where $T(E) = 1$; this arises because an infinite periodic array of molecules is modeled, and the calculated electronic wavefunctions are eigenstates of the functionalized CNT. Thus, there is no scattering per say due to the presence of the molecules. (In this approach, a value $T(E) \neq 1$ would only arise from consideration of wavefunction mismatch and scattering at the contacts.).

## 3. Results and Discussions

*3.1. Thermal conversion of CNT networks*

An AFM image of alkylated CNTs spincast on a blank $SiO_2$/Si substrate is shown in Figure 1. Unlike CNTs without surface functionalization [18], the alkylated CNTs are readily soluble in chloroform and easily spincast into networks of controlled density on $SiO_2$/Si substrates. Figure 1 shows thermogravimetric analysis (TGA) of the lithium alkylated CNTs heated in argon with a temperature ramp rate of 10 degrees per minute. These results indicate that the CNTs de-alkylate primarily over the temperature range 125 °C > T > 500 °C, as indicated by the weight % change in the TGA, with the largest (derivative) change around 200

°C. The mass loss observed is due to desorption of alkyl groups as previously confirmed by mass spectrometry [18, 19]. Results similar to those in Figure 1 were obtained when the CNTs were alkylated using sodium. Raman spectra comparing the pristine (unfunctionalized) CNTs both to the as-functionalized CNTs and to the functionalized CNTs that have been annealed (de-functionalized) are compared in [19]. The Raman spectra show that by annealing the functionalized CNTs to 800 °C (i.e. de-functionalizing) results in the recovery of the pristine CNT behavior [19]. Similarly, Raman spectra presented here (see Supplementary Data Figure S1) confirms that annealing the functionalized CNTs to 500 °C also results in the recovery of pristine behavior.

CNT network transistors were fabricated by spincasting the functionalized CNTs dispersed in chloroform onto $SiO_2$/Si substrates with pre-patterned electrodes. The transfer characteristics of a representative network device as-deposited (i.e. not annealed) and for two different annealing temperatures is shown in Figure 2, with an AFM image of the device shown in the inset. The current in the as-deposited CNT network is low due to the heavy alkylation of both the metallic and semiconducting CNTs. This electrical behavior is exhibited after annealing in the temperature range 25 °C < $T$ < 125 °C, denoted "I" on the TGA plot of Fig. 1. This is consistent with previous reports in the literature showing current suppression as a result of surface functionalization [13, 14, 15] and will be discussed in detail later. After annealing the device at 160 °C, the *ON* current increases while the *OFF* current remains relatively unchanged and the network device exhibits semiconducting behavior with an $I_{ON}/I_{OFF}$ ratio of >100. This semiconducting behavior is observed after annealing in the temperature range 125 °C < $T$ < 210 °C, denoted "II" in Fig. 1, corresponding to the onset of mass loss with increasing temperature, with the greatest $I_{ON}/I_{OFF}$ observed near the middle of this temperature range at ~160 °C.

Upon annealing to higher temperatures, the *ON* current continues to increase, while the *OFF* current increases dramatically and the device becomes metallic with an $I_{ON}/I_{OFF}$ ratio of <10. This anneal temperature regime corresponding to metallic behavior, $T > 210$ °C, is indicated by "III" on the TGA plot, starting near the maximum change in weight % with temperature and continuing thereafter. Applying a larger gate bias does not result in a conductance decrease (i.e. the devices do not turn off), indicating that the network truly exhibits metallic behavior as opposed to merely a threshold voltage shift. The TGA and electrical results are generally in good agreement with minor differences attributed to differences in heating conditions.

Next, the behavior of individual alkylated metallic CNTs was measured to identify the impact of functionalization on their electrical behavior. The behavior of one such metallic CNT device is shown in Figure 3, showing transfer characteristics as a function of increasing anneal temperature. For the as-deposited (not annealed) CNT, the current is very low (~100pA), similar to the behavior observed in the as-deposited networks. The 100-160 °C annealed current is similar to the *OFF* current of the semiconducting network transfer characteristic shown in Figure 2, suggesting that the current through the alkylated metallic CNTs determines the *OFF* current. As the metallic single-CNT device is annealed to higher temperatures, the current increases with little change in the gate response. This suggests that 1) the alkylation quenches metallic CNT current by a conductivity reduction mechanism, and 2) the degree of alkylation, controlled by anneal temperature, determines the conductivity. These conclusions are consistent with the previous findings [13, 14, 15] where a similar *ON*-current suppression mechanism was observed, with the degree of current suppression proportional to the extent of surface functionalization. However, in these previous works, the extent of surface functionalization was controlled by the concentration of diazonium regent applied to the network post-assembly, while here it is controlled by the extent of the initial alkylation and subsequent desorption upon heating. Because both approaches lead to similar behavior, the alkyl group molecules at CNT-CNT junctions or CNT-metal contacts likely do not play a major role in governing the network electrical behavior.

*3.2. Ab-initio modeling of the functionalized CNTs*

We used Density Functional Theory (DFT) calculations to support the hypothesis that functionalization reduces the conductivity. Electronic structure calculations were carried out for a variety of alkylated configurations and spin polarizations (14 different configurations in total), chosen based on the lowest-energy structures of covalently-functionalized CNTs investigated in recent studies [22, 23]. Our calculations indicate that configurations corresponding to alkyl adsorbates located at different sublattices give the lowest energy for both the (6,6) and (10,0) nanotube. For these particular lowest-energy configurations, all the nonmagnetic, ferrimagnetic, and anti-ferromagnetic solutions are nearly degenerate, with a small energy difference among them. We obtain binding energies of 0.90 and 0.94 eV for the alkylated (6,6) and (10,0) nanotubes, respectively, at a coverage of one alkyl group per 6 carbon rings. Doubling the density increases the binding energy to 1.46 eV and 1.53 eV for the (6,6) and (10,0) tubes respectively. This suggests that molecule-molecule interactions can play an important role in these systems.

The density of states for pristine and alkylated semiconducting (10,0) and metallic (6,6) CNTs are shown in Figures 4a-b with atomic models shown in the inset. The density of states of both systems is significantly distorted by the presence of the molecules, which can have an important impact on the conductance. Figures 4c-d show the calculated conductance of the pristine and alkylated CNTs as a function of energy. In good agreement with the experimental results, the conductance at the Fermi energy is reduced in the alkylated metallic CNT. In contrast to the metallic CNT, the semiconducting CNT conductance is less affected by the alkylation, with the conductance either slightly increasing or decreasing depending upon energy. This is consistent with the literature, which generally reports that greater surface functionalization is required to suppress the *ON* current of semiconducting CNTs [14]. We expect that a higher density of functionalization and random spatial distribution of molecules would further reduce the conductivity, leading to the low conductivity observed for the as-deposited devices, and the lack of an observable gate dependence. In addition, the networks typically require low temperature annealing (125-210 °C) to exhibit reasonable *ON* currents and semiconducting behavior, suggesting that the initial alkylation is heavy enough to reduce the semiconducting CNT *ON* current as well.

*3.3. Modeling of the thermal desorption*

The conversion of the CNT network devices from poorly conducting to semiconducting and then to metallic could arise if the semiconducting CNTs start to de-alkylate at lower temperatures compared to the metallic CNTs, or if the metallic CNTs are more heavily functionalized initially. To discriminate between the two mechanisms, we consider the TGA data of Figure 1 in more detail. Figure 5a shows the derivative of the mass fraction versus temperature obtained from Figure 1, indicating that desorption in this temperature range occurs mainly at temperatures of 200 °C and 400 °C. Previous measurements [18, 19] on this type of functionalized nanotubes have shown that both of these peaks correspond to mass loss from alkyl chains.

To model the TGA data, we assume first order kinetics for the coverage of molecules $\theta$ on the nanotube surface

$$\frac{d\theta}{dt} = -\nu e^{-\frac{E_{des}}{kT}} \theta \qquad (1)$$

where $t$ is time, $\nu$ is the attempt frequency, $T$ is the temperature, and $E_{des}$ is the desorption energy. We focus on the portion of the TGA curve obtained under a constant temperature ramp rate, i.e. $T(t) = T_0 + \gamma t$, and solve Eq. (1) numerically. The time dependence of the surface coverage then leads to the time dependence of the total mass of the sample, $m$,

$$m(t) = m_f \left[ 1 + \frac{m_{mol}}{m_C} \theta(t) \right] \qquad (2)$$

where $m_f$ is the final mass after complete molecular desorption, $m_{mol}$ is the mass of a molecule and $m_C$ is the mass of a carbon atom. The initial and final masses allow the determination of the initial surface coverage from the relation $\theta_0 = (m_0/m_f - 1)(m_C/m_{mol})$. Applied to the data of Figure 1, this gives an initial surface coverage of about one molecule per 10 carbon atoms.

We first considered two *single* desorption energies to represent the two peaks of Figure 5a. While it is possible to reproduce the peak positions with single desorption energies, it is not possible to reproduce the width of the peaks, even if the attempt frequency and $E_{des}$ are varied over a large range. (We have also tried zeroth and second order kinetics, both of which also fail to describe the data). However, good agreement with the data can be obtained if a tri-modal Gaussian distribution of desorption barriers is considered: the solid line in Figure 5a is obtained with three distributions with average energies of 1.3 eV, 1.42 eV, and 1.9 eV, and widths of 0.06 eV, 0.12 eV, and 0.2 eV, respectively (Figure 5b), and an attempt frequency of 1.2 THz [24]. (These distributions of desorption energies could arise due to variations in the binding configuration of molecules, interactions with other molecules, binding with defects, etc.). The two distributions at 1.3eV and 1.42 eV are needed to reproduce the asymmetric shape of the 200 °C peak. The three peaks correspond to total integrated coverages of 12%, 19%, and 16% of the total carbon atoms. It should be noted that the values for the average energies and the distribution of energies for peaks 1 and 2 agree well with the range of values obtained from the ab initio calculations when density effects are considered.

To relate these results to the network FET annealing experiments, the coverage from each of the three peaks was calculated after annealing for 20 minutes at constant temperature, using the parameters extracted from the TGA fit. Figure 5c shows that over the temperature range of region II, no desorption occurs corresponding to peak 3, indicating that this desorption process does not play a role in converting the network electrical properties. In contrast, the coverages corresponding to peaks 1 and 2 change significantly between 100 °C and 200 °C, with the coverage corresponding to peak 2 starting at a larger value and dropping at a lower rate than peak 1. These results correlate well with the electrical data, and suggest that peaks 1 and 2 are due to the functionalization of the semiconducting and metallic CNTs, respectively. Taken together, the experimental and modeling results point to a higher initial functionalization of the metallic CNTs, and a preferential thermal desorption of the semiconducting CNTs in region II.

*3.4. Physical removal of metallic CNTs*

While the above results indicate that annealing can be used to control the degree of functionalization and the electrical behavior of the CNT network devices, a more robust approach is one where the metallic CNTs can be *physically* removed from the networks. We demonstrate such an approach by utilizing alkylation and annealing in combination with the fact that more heavily alkylated CNTs have greater solubility in chloroform, to remove metallic CNTs directly from network devices. As previously discussed, the metallic CNTs retain or partially retain their surface functionalization to higher anneal temperatures compared to the semiconducting CNTs. Therefore, by annealing to a temperature just high enough to de-alkylate the semiconducting CNTs, while leaving the metallic CNTs partially alkylated, enables the use of a chloroform bath and rinse to remove the metallic CNTs. Figure 6a shows a schematic of this annealing and chloroform rinse process to eliminate the metallic CNTs. Figure 6b shows an example alkylated CNT network device of similar density to that shown in Figure 2 that was first annealed to 210 °C, then bathed and rinsed in chloroform. The absence of gate modulation but still relatively low *ON* current after the annealing step indicates that the metallic CNTs have partially de-alkylated at this temperature. Upon rinsing with chloroform, a relatively small drop in *ON* current is observed, accompanied with a significant drop in *OFF* current. The large drop in the *OFF* current indicates that most of the metallic CNTs were removed from the network, due to remaining alkylation that makes them soluble, and thus removable, in chloroform. The result is that the CNT network with an initial $I_{ON}/I_{OFF}$ ratio <10 is turned semiconducting with an $I_{ON}/I_{OFF}$ ratio >$10^3$ through the physical removal of metallic CNTs. The success rate of this technique is ~81% (17 of 21 devices tested for the conditions described above). It should also be noted that the *OFF* current is lower compared to the semiconducting network presented in Figure 2 after 160 °C annealing alone, where the metallic CNTs were still present. This is further evidence that the metallic CNTs were effectively removed from the network with the chloroform rinse. Additionally, the semiconducting behavior of the network is preserved with higher temperature annealing, again indicating that the metallic CNTs have been removed. It should also be noted that as higher temperature anneals are used in the first step, it is not possible to obtain the semiconducting behavior of the network with the chloroform rinse. This is consistent with de-alkylation of the metallic CNTs at higher temperature, making them insoluble in chloroform.

To verify that CNT material was removed from the CNT network devices, an analysis of AFM images taken on low density networks before and after the annealing and chloroform rinse was performed. Figure 6c shows an AFM image of a low density of alkylated CNTs dispersed on $SiO_2$/Si, while Figure 6d shows a similar image at a nearby position after annealing to 210 °C and following a chloroform rinse. Analysis of these images indicate that there are ~825 CNTs in a 10μm x 10μm area (~8.25 CNTs/μm$^2$) before the anneal and rinse (Figure 6c) compared to ~433 CNTs in a 10μm x 10μm area (4.33 CNTs/μm$^2$) after the anneal and rinse (Figure 6d). This reduction in CNT density was found in all samples with similar initial CNT density that had been analyzed, demonstrating that CNTs are being physically removed from the devices. The analysis of networks with higher densities, corresponding to that of Figure 2 and 6b, resulted in an undercounting of CNTs, so only lower limits to the density could be measured. Nonetheless, these lower limits still indicate a reduction in CNT material following the anneal and chloroform rinse.

Finally, experiments were performed to eliminate the possibility that the increased $I_{ON}/I_{OFF}$ ratio in the annealed and chloroform rinsed networks was due to the non-selective removal of CNT material and, therefore, reduction in connecting metallic pathways through the network due to an overall decrease in the network density. CNT network devices with a lower CNT density compared to the post-rinsed network devices used to acquire the data in Figure 6b were fabricated. Upon annealing, these lower density network devices still became metallic (See

Supplementary Data Figure S2), indicating that this lower network density, and therefore the densities of the devices corresponding to Figure 6b, is still above the percolation threshold density required for a sufficient number of metallic pathways to exist between electrodes. It can be concluded that the increase in $I_{ON}/I_{OFF}$ ratio following the chloroform rinse is indeed due to the selective elimination of metallic CNTs from the network.

## 4. Summary and conclusions

In conclusion, a method utilizing alkylation of CNTs, which makes the CNTs easily dispersible on substrates, followed by annealing is demonstrated as a means to control the semiconducting to metallic electrical behavior of CNT network devices. The alkylation and annealing is combined with a chloroform treatment to physically remove the metallic CNTs from the networks directly on $SiO_2$/Si substrates. This work presents an alternative method to improving the performance of CNT array and network FETs by an "on-substrate" technique to remove metallic CNTs.

**Acknowledgements**

This project is supported by the U. S. Department of Energy, Office of Science, through the National Institute for NanoEngineering (NINE) at Sandia National Laboratories, a multiprogram laboratory operated by Sandia Corporation, a Lockheed Martin Company, for the United States Department of Energy under contract DE-AC04-94-AL85000. W. E. B is supported by the Welch Foundation (C-0490). R. H. Hauge acknowledges support from the Lockheed Martin Lancer program.

**Figure Captions**

Figure 1. The main panel shows TGA data for the lithium alkylated CNTs. Regions I, II, and III correspond to temperature ranges where low conductivity (I), semiconducting behavior (II), and metallic behavior (III) are observed in CNT network devices. The inset shows an AFM amplitude image (10μm x 10μm) of alkylated CNTs on a $SiO_2$/Si substrate.

Figure 2. Transfer characteristics measured at $V_{ds}$=0.1V for a representative CNT network FET (electrode spacing ~ 1.2 μm), as-deposited (no anneal) and after annealing at temperatures of 160 and 420 °C. The inset shows an AFM amplitude image of the device (2μm x 2μm).

Figure 3. Transfer characteristics measured at $V_{ds}$=0.1V for a single metallic CNT FET (channel length $L$ ~ 850nm), for increasing anneal temperatures, from bottom to top.

Figure 4. Calculated properties of single-wall CNTs functionalized with alkyls. Panels (a) and (b) compare the density of states for the (10,0) semiconducting and (6,6) metallic CNTs before and after functionalization, while panels (c) and (d) show the resulting conductance.

Figure 5. Modeling of the TGA experiments. Panel (a) shows experimental data for the derivative of the mass as a function of temperature, obtained from Figure 1 (solid squares). The solid line shows a fit obtained by solving Eq. (1) using the distribution of energies shown in panel (b). Panel (c) shows the simulated coverage due to each of the three peaks of panel (b) during a constant temperature anneal for 20 minutes. The shaded area is the temperature range corresponding to region II of Figure 1.

Figure 6. Chloroform rinse experiments. (a) A schematic of the annealing and chloroform rinse process showing the removal of the metallic CNTs. In the first step, a functionalized CNT network device is fabricated. In the second step, annealing at low temperatures leads to the preferential de-alkylation of the semiconducting CNTs. In the final step, a chloroform rinse is used to preferentially remove the metallic CNTs. (b) Transfer characteristics measured at $V_{ds}$=0.1V for a CNT network FET (electrode spacing ~ 1.1 µm), (1) after annealing at low temperature (210 °C, top curve) and (2) subsequent rinsing in chloroform (bottom curve). (c,d) AFM images (10µm x 10µm) with a CNT trace mask generated with the SIMAGIS software showing (c) the alkylated CNTs as-deposited on $SiO_2$/Si, and (d) after annealing to 210 °C and rinsing in chloroform. The AFM images in (c) and (d) were taken at nearby positions on the same wafer. A lower density of CNTs compared to the networks in Figures 2 and 6b is shown to more clearly illustrate the removal of CNT material, as well as for the purposes of more accurate image analysis.

# Physical Removal of Metallic Carbon Nanotubes from Nanotube Network Devices Using a Thermal and Fluidic Process

Alexandra C. Ford,[1] Michael Shaughnessy,[1] Bryan M. Wong,[1] Alexander A. Kane,[1] Oleksandr V. Kuznetsov,[2] Karen L. Krafcik,[1] W. E. Billups,[2] Robert H. Hauge,[2] François Léonard[1*]

[1]Sandia National Laboratories, Livermore, California 94551, United States

[2]Department of Chemistry, Rice University, Houston, Texas 77005, United States

*fleonar@sandia.gov

**Supplementary Data**

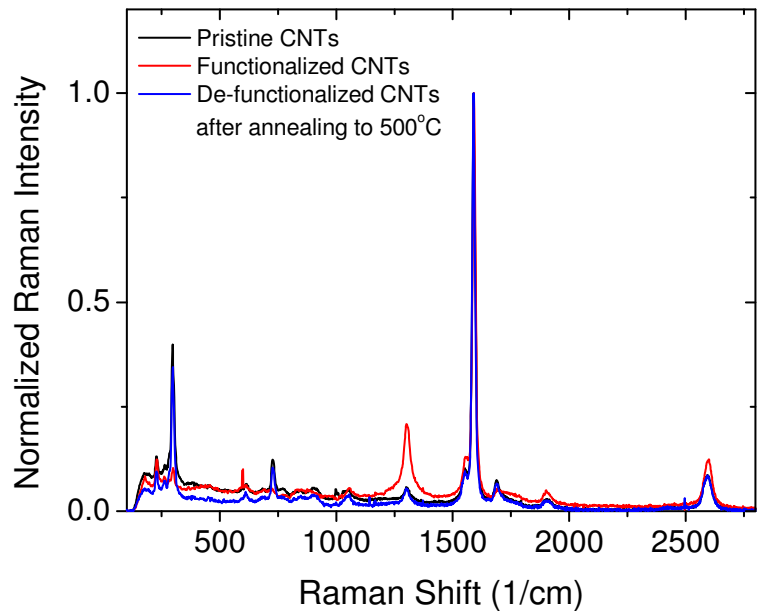

**Figure S1.** Comparison of Raman spectra for pristine CNTs, functionalized CNTs, and de-functionalized CNTs after annealing to 500 $^{o}$C. The Raman shows that by annealing the functionalized CNTs to 500 $^{o}$C (i.e. de-functionalizing) results in the recovery of the pristine CNT behavior.

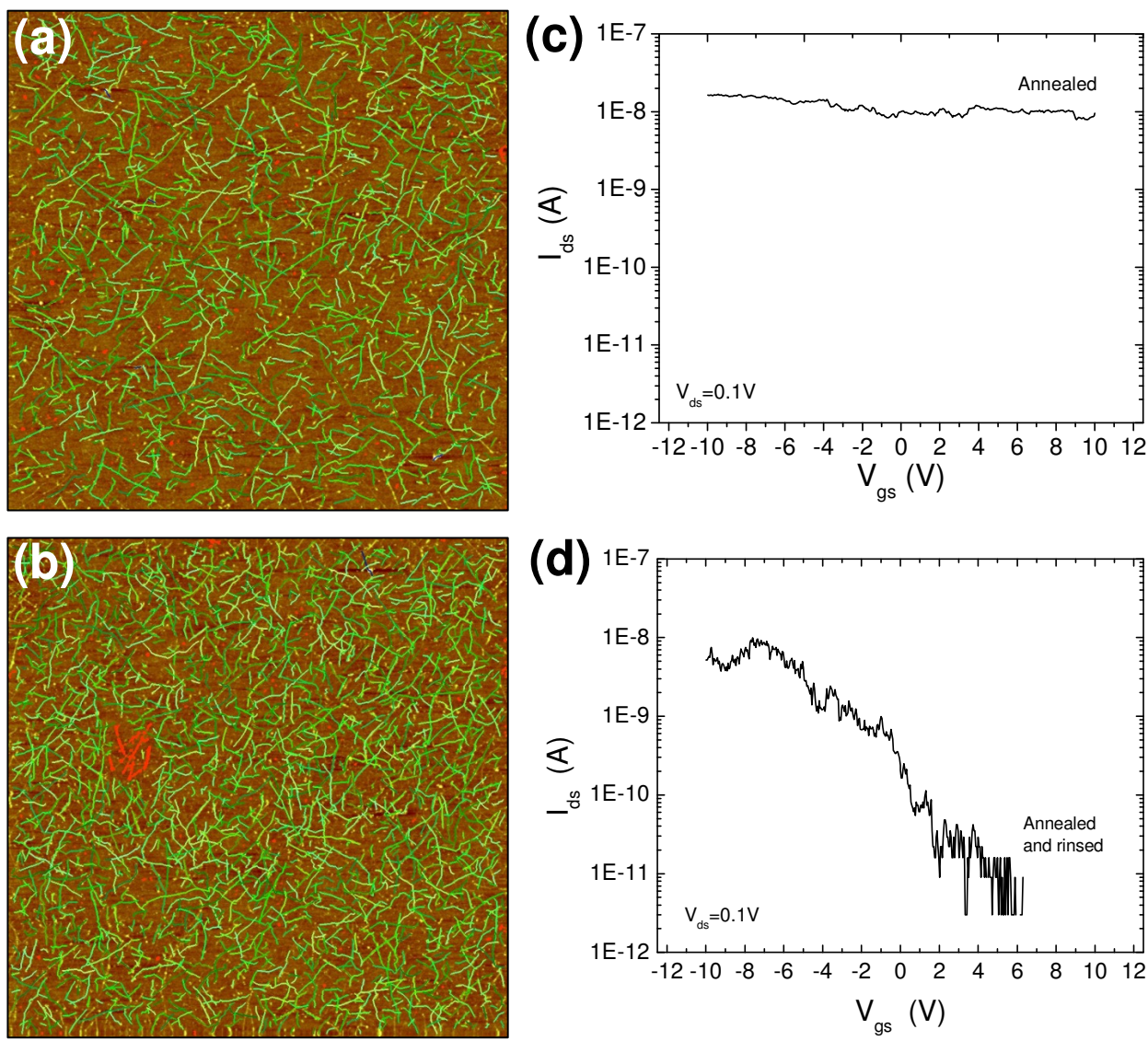

**Figure S2.** Lower density networks compared to the annealed and chloroform rinsed networks. AFM images (10μm x 10μm) analyzed using SIMAGIS with CNT traces shown for clarity for (a) lower density networks (~12 CNTs/μm$^2$, accurately counted using SIMAGIS) and (b) annealed and chloroform rinsed networks (>16 CNTs/μm$^2$, a lower limit given some underestimation of CNTs counted). (c) Transfer characteristics for a lower density network device at $V_{ds}$=0.1V (electrode spacing ~1μm), showing metallic behavior after annealing, confirming that metallic pathways connect electrodes at this density. (d) Transfer characteristics for the annealed and chloroform rinsed network device (reproduced from Figure 6b for comparison here). The network density (b) of this device is clearly above the density (a) resulting in metallic pathways, implying that metallic CNTs have been removed by the chloroform rinse.

**Figure 1**

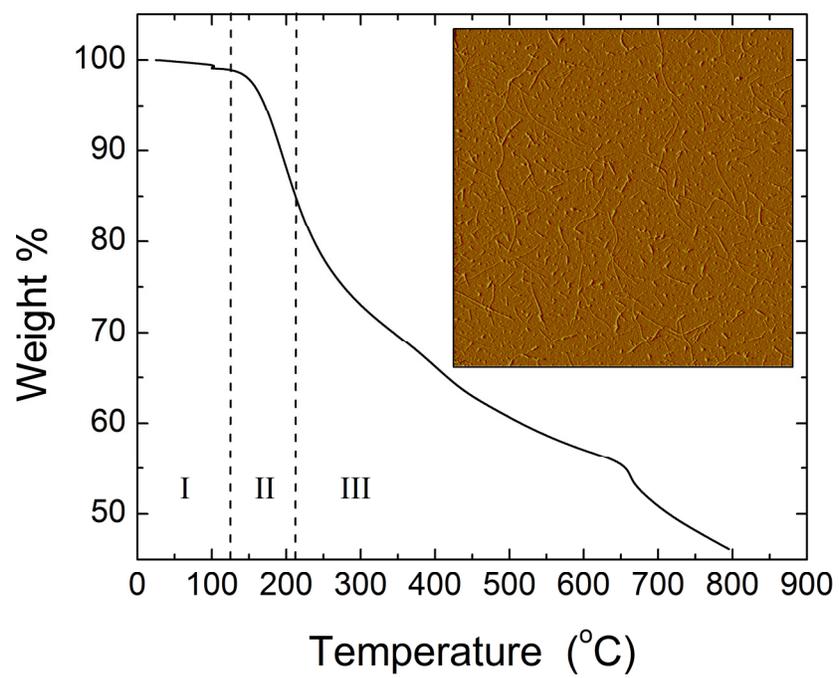

**Figure 2**

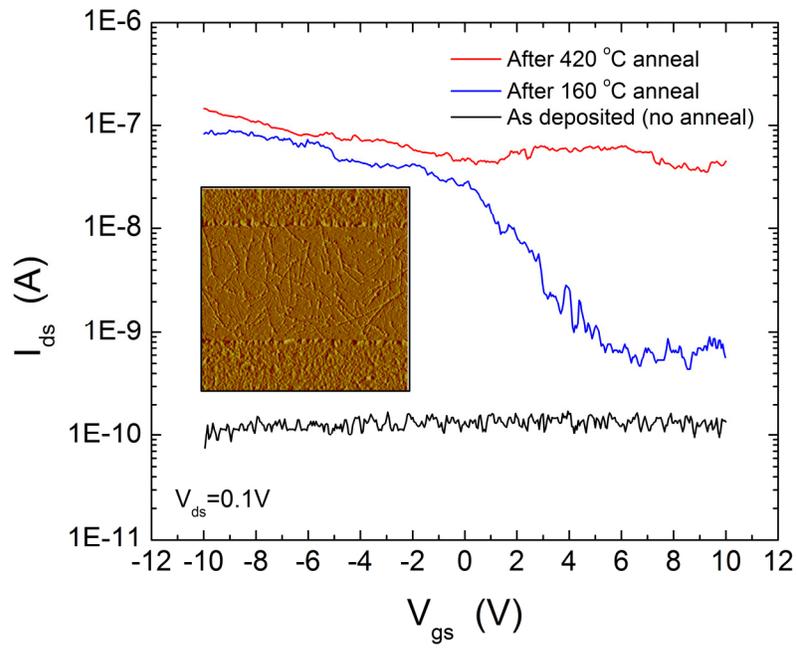

**Figure 3**

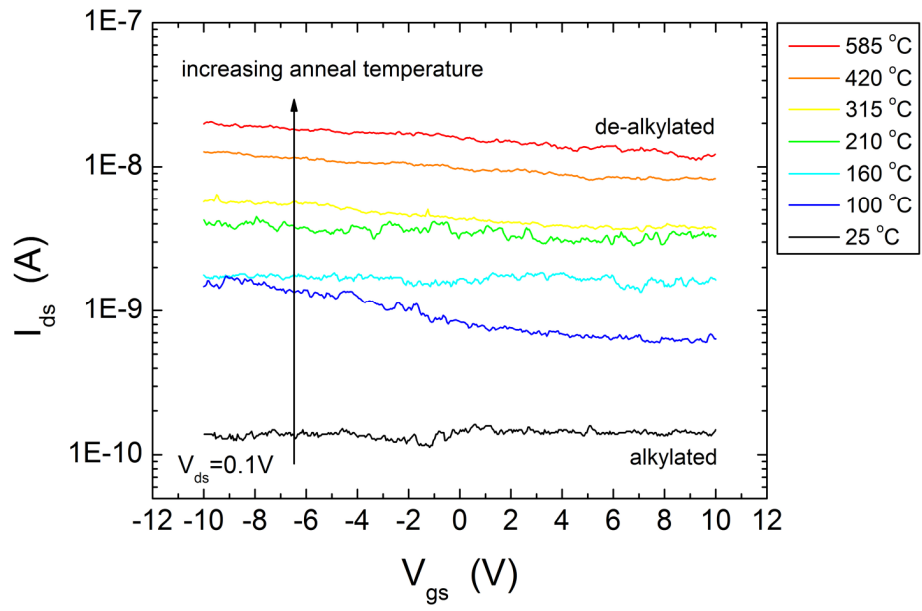

**Figure 4**

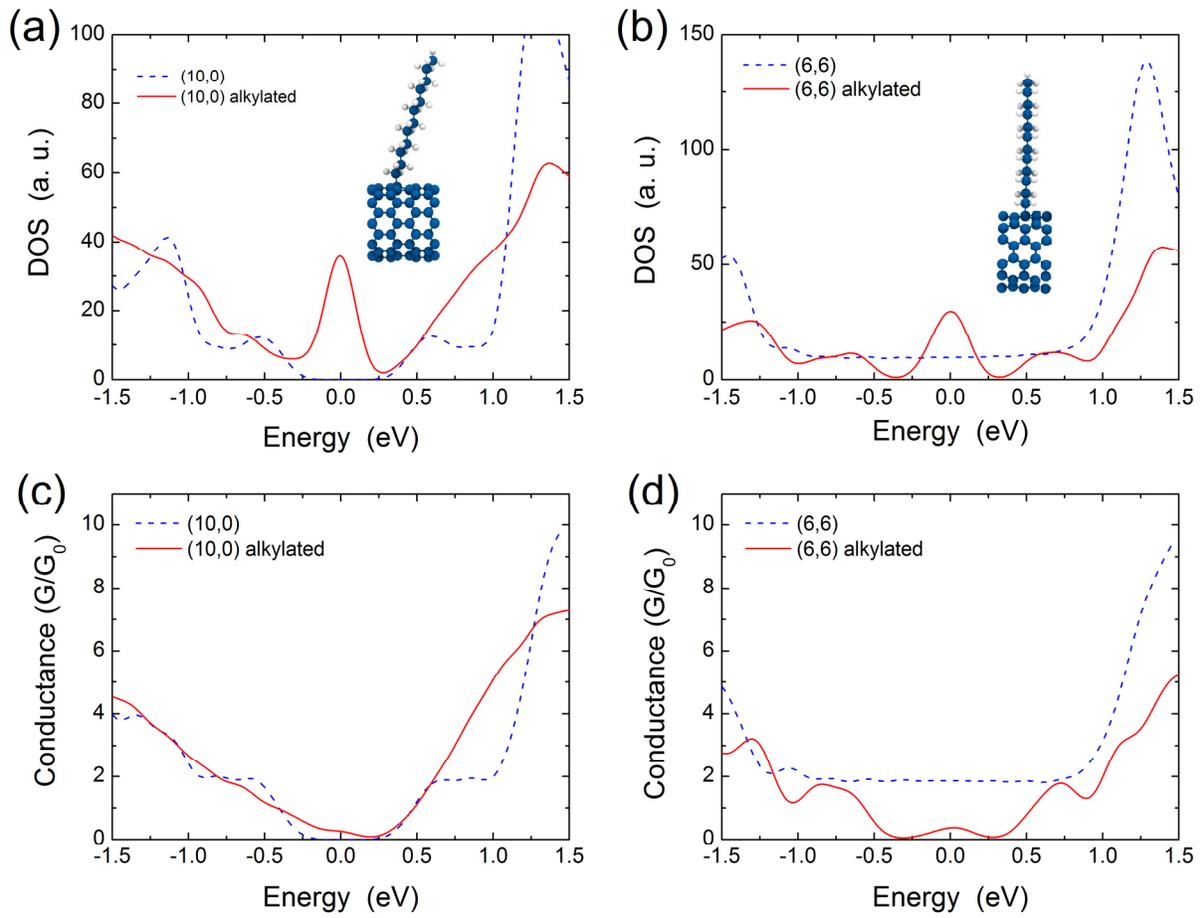

**Figure 5**

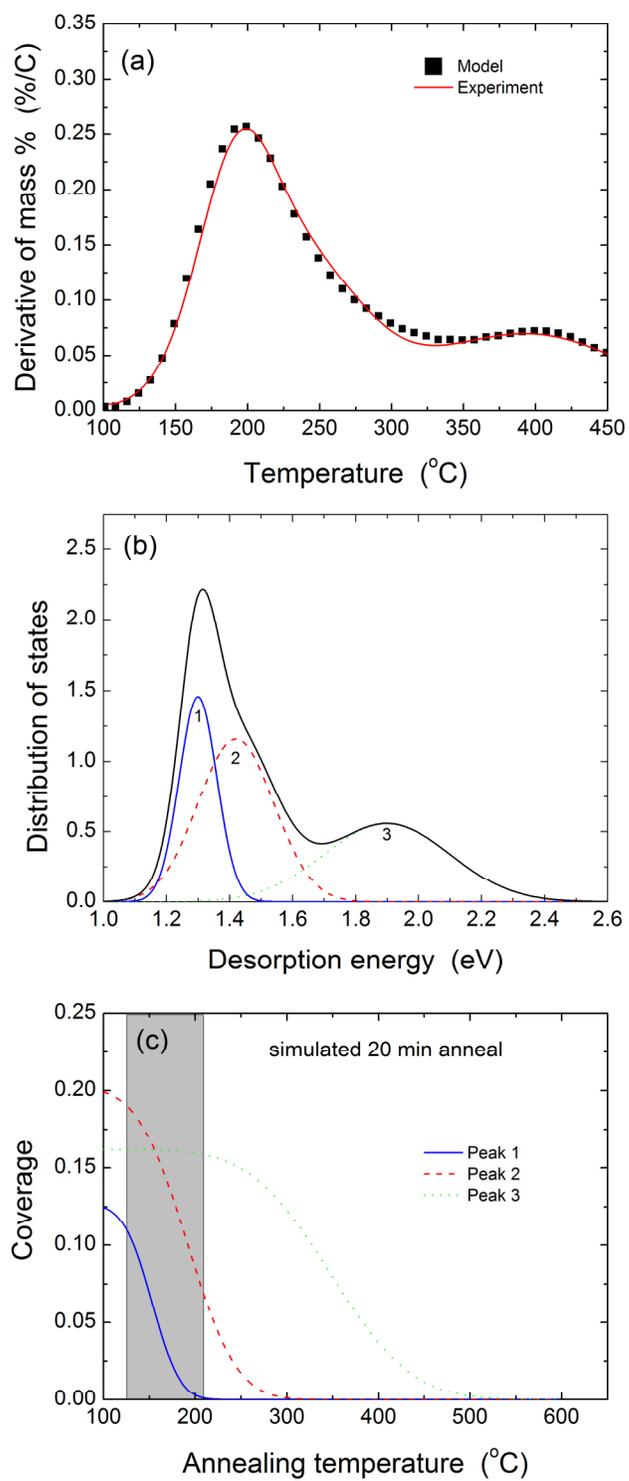

**Figure 6**

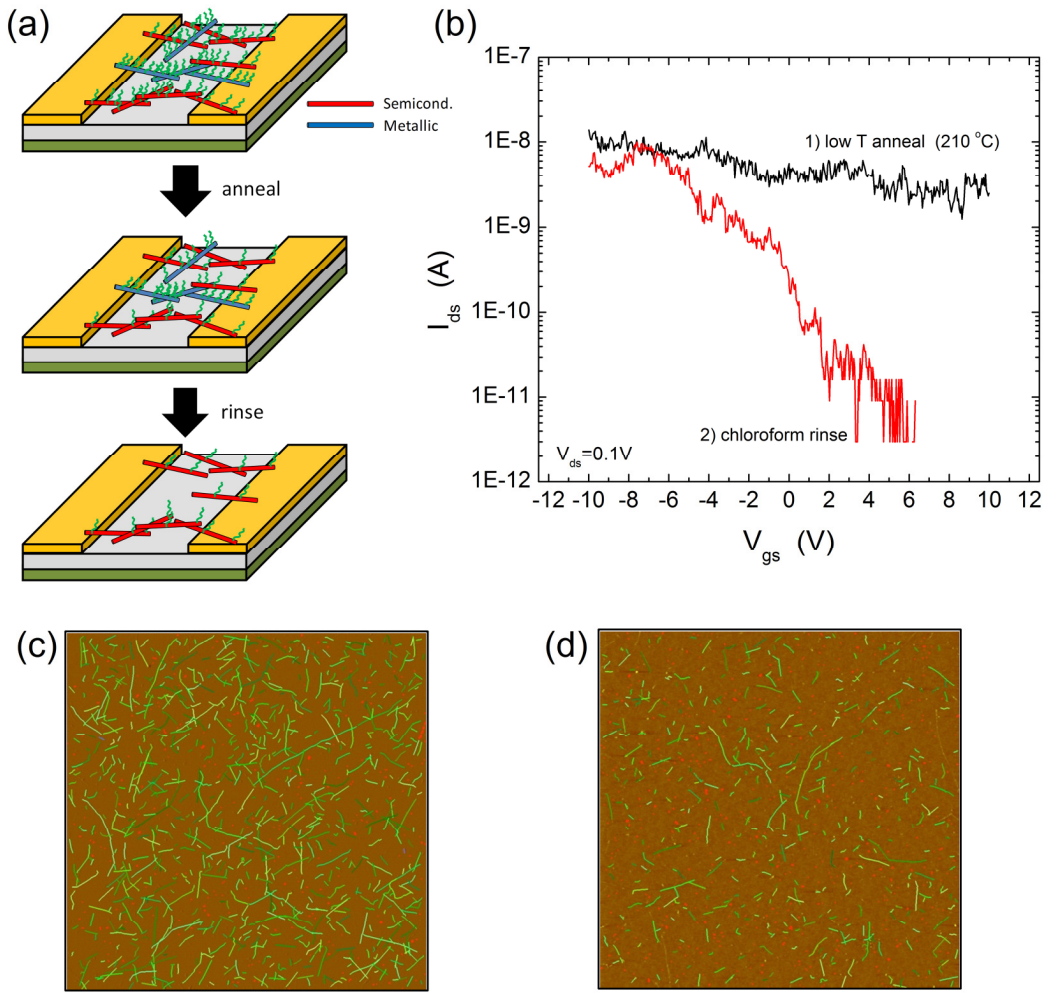